\documentclass[10pt]{article}
\usepackage[a4paper,margin=1in]{geometry}
\usepackage{amsmath}
\usepackage{graphicx}
\usepackage{indentfirst}
\usepackage{hyperref}
\usepackage{float}
\usepackage{booktabs}
\usepackage{authblk}

\title{Integrative Analysis of Epigenetic, Transcriptomic, and Metabolomic Responses to Arsenic Exposure Using Coupled Matrix Factorization}

\author[1]{Sujit Silas Armstrong Suthahar}
\author[1,*]{Patrick Allard}

\affil[1]{University of California, Los Angeles, Los Angeles, CA 90095, USA}

\date{}

\begin{document}
\maketitle

\section{Introduction and Motivation}
Arsenic (As), a naturally occurring element, poses significant toxicity risks in its inorganic form (iAs). It is a pervasive environmental toxin associated with severe health implications, including cancer, cardiovascular disease, and endocrine disruption [1]. Its global distribution and bioaccumulation make it a critical public health concern, particularly relevant to computational toxicology and environmental health assessment. While extensive research has been conducted to study arsenic-induced toxicity, the mechanisms by which it influences molecular pathways remain poorly understood. This complexity arises from arsenic's ability to induce widespread epigenetic reprogramming, alter gene expression, and disrupt metabolic homeostasis through interconnected regulatory networks [2,3].

Despite advances in high-throughput techniques, existing studies often analyze epigenomic, transcriptomic, and metabolomic data independently. This approach overlooks the intricate interplay among these layers of regulation. Integrating such multi-omics data provides an opportunity to uncover novel regulatory mechanisms underlying arsenic toxicity. Recent advances in computational biology have demonstrated the power of tensor factorization and coupled matrix factorization methods for joint multi-omics analysis [4,5,6]. Notably, Multi-Omics Factor Analysis (MOFA) has emerged as a framework for unsupervised integration of multi-omics datasets [7], while PARAFAC2-based methods have shown particular promise for handling temporally irregular and heterogeneous data structures [8,9].

Previous studies have explored similar computational approaches for integrative data analysis but have not systematically applied them in toxicological contexts. Bagherian et al. developed coupled matrix–matrix and coupled tensor–matrix completion methods to predict drug–target interactions [10]. Similarly, Erbe et al. applied the CoGAPS matrix factorization algorithm and transfer learning with projectR to integrate single-cell ATAC-seq datasets [11]. Duren et al. leveraged coupled non-negative matrix factorization for joint analysis of scRNA-seq and scATAC-seq data [12]. Recent developments include PARAFAC2-RISE for single-cell experiments across conditions [13], methPLIER for DNA methylation analysis using non-negative matrix factorization with biological pathway constraints [14], and federated PARAFAC2 tensor factorization for computational phenotyping [15]. However, these approaches have not been systematically applied to joint analysis of DNA methylation, gene expression, and metabolomics in environmental toxicology. In our study, we apply CMF using the parafac2\_aoadmm model to integrate and analyze RRBS, RNA-seq, and metabolomics data from mouse ESCs and EpiLCs treated with arsenic [16].

\section{Problem Definition}
This study aims to investigate how arsenic exposure influences molecular pathways across the epigenome, transcriptome, and metabolome. Specifically, we seek to uncover general trends in global methylation dysregulation that drive transcriptional and metabolic changes in ESCs and EpiLCs. Additionally, we aim to identify distinct regulatory networks and mechanisms underlying arsenic toxicity for applications in computational toxicology and risk assessment.

To address this as a data analytics problem, we employed CMF, which enables the joint factorization of datasets sharing common features (columns) but differing in dimensions (row counts). This approach can model arsenic's effects as a coupled decomposition problem, where each dataset can provide unique yet complementary insights. Unlike traditional non-negative matrix factorization approaches that assume data non-negativity, our coupled matrix factorization approach allows for negative values, enabling more comprehensive modeling of bidirectional regulatory changes [17]. The broader objective is to link these molecular alterations to the pathophysiological outcomes associated with arsenic-related diseases, including cancer and developmental disorders, while advancing computational toxicology methodologies for environmental health assessment.

\section{Methods}

\subsection{Algorithm Description}
The analysis was conducted using the \texttt{parafac2\_aoadmm} model, a specialized implementation of Coupled Matrix Factorization (CMF) [4]. CMF is designed to jointly factorize multiple matrices that share column features but differ in row dimensions, making it particularly suitable for multi-omics integration where different molecular layers have varying numbers of features. Unlike conventional tensor decomposition techniques that assume alignment across dimensions, PARAFAC2 addresses the challenge that multi-omics technologies do not measure the same molecular entities across all modalities [18]. The model follows the formulation:  
\[
X^{(i)} \approx B^{(i)}D^{(i)}C^T
\]  
where \( X^{(i)} \) represents the \( i \)-th input matrix (e.g., RRBS, RNA-seq, or metabolomics), \( B^{(i)} \) is a collection of factor matrices for each dataset capturing sample-specific patterns, \( D^{(i)} \) consists of diagonal matrices indicating the signal strength of each dataset, and \( C \) is a shared factor matrix representing common features across the datasets.

To achieve this decomposition, PARAFAC2 relies on alternating optimization combined with the alternating direction method of multipliers (AO-ADMM) [16]. To fit a coupled matrix factorization for integrating \textbf{RRBS}, \textbf{RNA-seq}, and \textbf{metabolomics} data, we solve the following optimization problem:

\[
\min_{\mathbf{A}, \{\mathbf{B}^{(i)}\}_{i=1}^3, \mathbf{C}} \frac{1}{2} \sum_{i=1}^3 \frac{\|\mathbf{B}^{(i)} \mathbf{D}^{(i)} \mathbf{C}^\top - \mathbf{X}^{(i)}\|^2}{\|\mathbf{X}^{(i)}\|^2},
\]
where \(\mathbf{A}\) is the matrix constructed by stacking the diagonal entries of all \(\mathbf{D}^{(i)}\)-matrices (\(i = 1, 2, 3\) corresponding to \textbf{RRBS}, \textbf{RNA-seq}, and \textbf{metabolomics} data, respectively). However, this optimization problem does not yield a unique solution, as fitting coupled matrix factorization can produce different factor matrices. This makes it challenging to interpret the factor matrices directly.

To address this, we introduce regularization terms, leading to the following revised optimization problem:

\[
\min_{\mathbf{A}, \{\mathbf{B}^{(i)}\}_{i=1}^3, \mathbf{C}} \frac{1}{2} \sum_{i=1}^3 \frac{\|\mathbf{B}^{(i)} \mathbf{D}^{(i)} \mathbf{C}^\top - \mathbf{X}^{(i)}\|^2}{\|\mathbf{X}^{(i)}\|^2} 
+ \sum_{n=1}^{N_A} g_n^{(A)} (\mathbf{A}) 
+ \sum_{n=1}^{N_B} g_n^{(B)} (\{\mathbf{B}^{(i)}\}_{i=1}^3) 
+ \sum_{n=1}^{N_C} g_n^{(C)} (\mathbf{C}),
\]
where the \(g\)-functions represent regularization penalties, and \(N_A\), \(N_B\), and \(N_C\) denote the number of regularization terms applied to \(\mathbf{A}\), \(\{\mathbf{B}^{(i)}\}_{i=1}^3\), and \(\mathbf{C}\), respectively.

\subsection{Software Implementation}
We implemented the parafac2\_aoadmm model using the MatCouply library in Python, which supports flexible CMF implementations [4]. The existing implementation provided by MatCouply was customized to handle the decomposition of RRBS, RNA-seq, and metabolomics data. Additional visualization modules were developed to interpret the factor matrices ($B^{(i)}, D^{(i)}, C$), enabling insights into the underlying patterns of the data. Additionally, the L1 norm was enforced to introduce a sparsity constraint to help with interpretability by reducing the complexity of the model. Non-negativity constraints were explicitly disabled to explore a broader solution space and capture bidirectional regulatory changes characteristic of biological systems. The maximum number of iterations was set to 100, ensuring adequate computational time for the algorithm to converge to a stable solution. The complete code implementation has been uploaded to GitHub (\url{https://github.com/sujitsilas/cmf-multiomic-analysis/})for reproducibility purposes.

\subsection{Data and Preprocessing}
Each dataset was normalized according to its specific requirements. For the RRBS data, beta values (methylation values normalized for coverage) were obtained following best practices for DNA methylation analysis. To enhance biological interpretability and extract more meaningful signals from the methylation dataset, 414,234 CpG sites were annotated based on ChromHMM inferred chromatin state annotations from mouse embryonic stem cells (mm10 genome) [19]. The ChromHMM annotations categorize genomic regions into biologically relevant chromatin states including active promoters, enhancers, transcribed regions, and repressed chromatin domains. Following annotation, beta values were aggregated based on 12 candidate chromatin state annotations, reducing dimensionality while preserving biologically meaningful methylation patterns associated with distinct regulatory elements. RNA-seq data was processed to produce transcripts per million (TPM) normalized counts, and the metabolomics data was normalized using total ion count. Rows in each dataset represent molecular entities (e.g., chromatin state-aggregated methylation values, genes, or metabolites), while columns correspond to experimental conditions or replicates. Additionally, genes with low expression levels (sum of TPM >10 across all samples) and genomic regions with low-coverage methylation (sum of beta values > 6 across all samples) were filtered out to reduce analytical noise and improve model stability.

\begin{enumerate}
    \item \textbf{RRBS (12 × 12)}: Profiles DNA methylation patterns in ESCs and EpiLCs under arsenic treatment using 12 candidate chromatin state annotations, providing comprehensive epigenetic coverage for toxicological analysis.
    \item \textbf{RNA-seq (26,210 × 12)}: Quantifies transcriptomic changes in response to arsenic exposure, capturing gene expression alterations across the genome.
    \item \textbf{Metabolomics (137 × 12)}: Measures metabolic fluxes in treated and control samples, enabling assessment of arsenic-induced metabolic perturbations.
\end{enumerate}

\section{Results}

\subsection{Model Optimization and Component Selection}
We divided the analysis in two parts where Figure 2 delineates the differences between ESCs and EpiLCs discerning changes between the cell type. Figure 3 highlights within cell type differences due to the toxic response to arsenic captured by the two component model (captures patterns in control ESCs) and three component model (captures patterns in ESCs treated with arsenic). In both the parts of the analysis, the three-component model provided the best balance between reconstruction accuracy and biological interpretability. As shown in Figure 2A, the fit score (reconstruction accuracy) plateaued around three components and remained stable through higher component numbers, while feasibility gaps (constraint satisfaction metrics) showed optimal performance at three components with increased instability at higher dimensions. Figure 3A shows independent model optimization results for ESCs specifically, confirming the robustness of the three-component solution across different experimental conditions.

\begin{figure}
    \centering
    \includegraphics[width=0.95\linewidth]{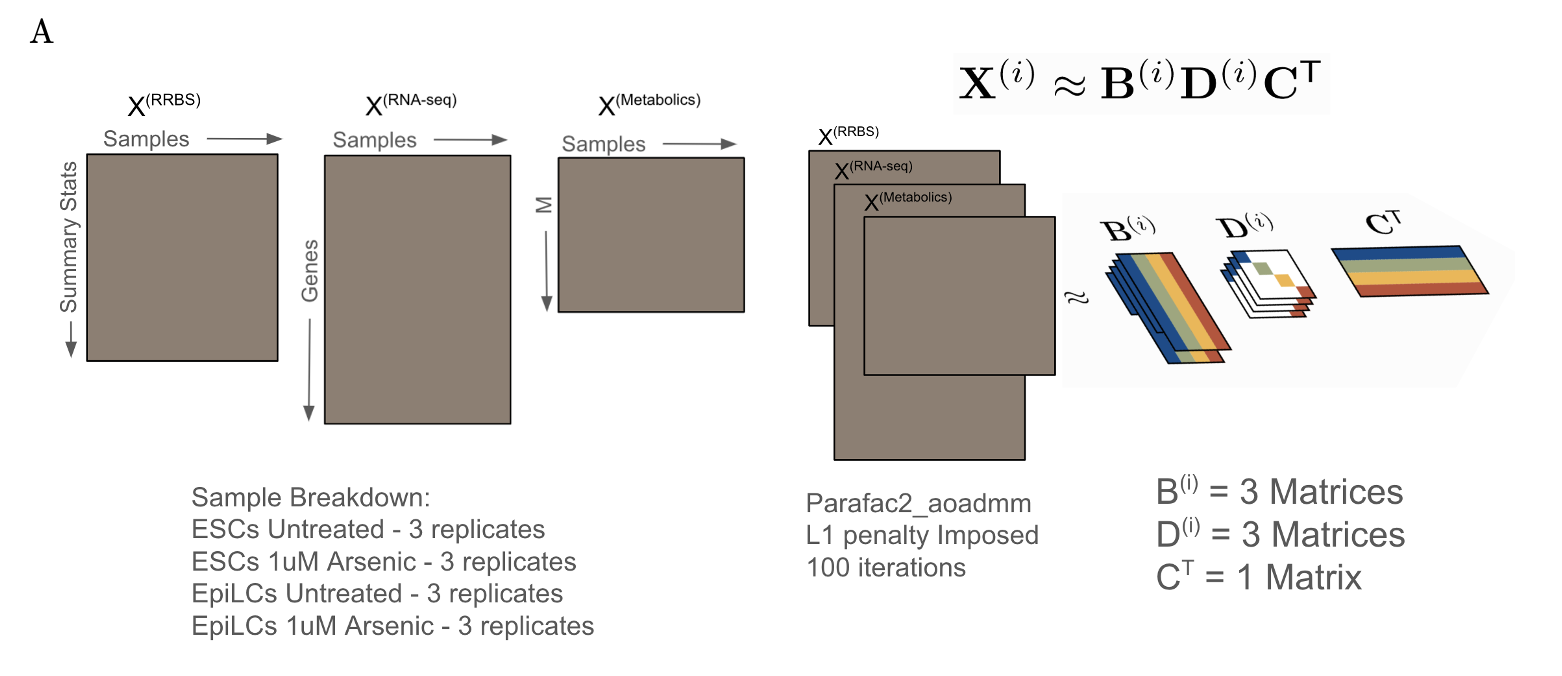}
      \caption{
        \textbf{Coupled matrix factorization framework and model optimization.} \textbf{A.} Schematic representation of the PARAFAC2-based coupled matrix factorization approach applied to integrate RRBS (12 chromatin state annotations), RNA-seq (26,210 genes), and metabolomics (137 metabolites) data from ESCs and EpiLCs under control and 1µM arsenic treatment conditions. The algorithm decomposes each data matrix $X^{(i)}$ into modality-specific factors $B^{(i)}$ and $D^{(i)}$, and a shared factor matrix $C^T$ representing common sample patterns across all three omics layers. This framework enables identification of coordinated molecular responses to arsenic exposure across epigenetic, transcriptomic, and metabolic levels.
    }
    \label{fig:model-framework}
\end{figure}

\subsection{Modality-Specific Contributions and Component Patterns}
Analysis of the component strength matrices ($\mathbf{D}^{(i)}$) revealed distinct contribution patterns across the three omics modalities. RRBS (DNA methylation) data dominated, while RNA-seq and metabolomics showed varying contributions across components. The shared factor matrix ($\mathbf{C}^T$) analysis across experimental conditions revealed biologically meaningful patterns (Figure 2C). Component 2 clearly distinguished between ESCs and EpiLCs regardless of arsenic treatment, capturing inherent epigenetic differences across the cell types delineating the differences between ESCs and EpiLCs. Components 2 and 3 captured arsenic-specific molecular responses in the within cell type comparison.

\begin{figure}
    \centering
    \includegraphics[width=0.95\linewidth]{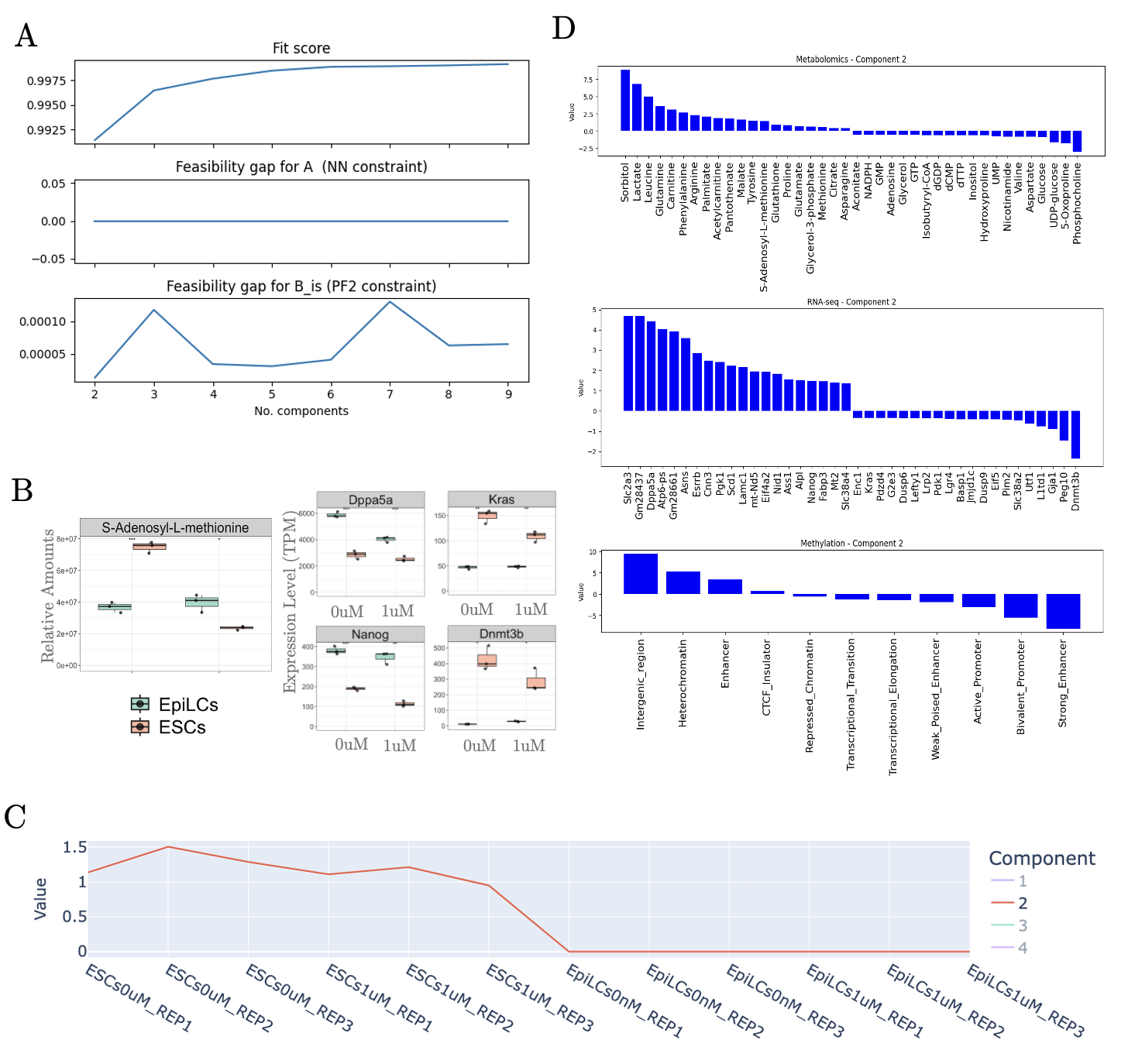}
        \caption{\textbf{Component based feature-level analysis identifies biologically relevant cell type specific molecular signatures.} \textbf{A.} Model fitting performance demonstrating reproducible three-component solution. \textbf{B.} Expression validation of key genes identified by the model: S-adenosyl-L-methionine levels, and expression of pluripotency markers (Dppa5a, Nanog) and methylation regulators (Kras, Dnmt3b). ESCs show higher Nanog expression and lower Dnmt3b compared to EpiLCs, with arsenic treatment modulating these patterns. Stars indicate the level of statistical significance based on a two-sample t-test. \textbf{C.} Shared factor analysis showing distinct component patterns across experimental conditions, with Component 2 separating cell types \textbf{D.} Feature-level factor loadings for each omics modality and component, demonstrating the biological relevance of identified molecular signatures.}
    \label{fig:component-analysis}
\end{figure}

\subsection{Molecular Feature Identification and Validation}
Analysis of the modality-specific factor matrices ($\mathbf{B}^{(i)}$) identified several biologically relevant molecular features (Figure 3B,D). Among the top-ranking metabolites, we identified significant loading for S-adenosyl-L-methionine and related compounds. Our analysis highlighted \textit{Gapdh} among the most variable genes across components. The analysis also identified \textit{Dnmt3b} as a key variable gene. Among metabolites, our analysis identified compounds related to glutathione metabolism, including 5-oxoproline (pyroglutamate). Additional molecular targets showing differential responses included ATP synthase subunit (Atp6-ps), heat shock protein (Hspa8), and thioredoxin-interacting protein (Txnip).

\begin{figure}
    \centering
    \includegraphics[width=0.95\linewidth]{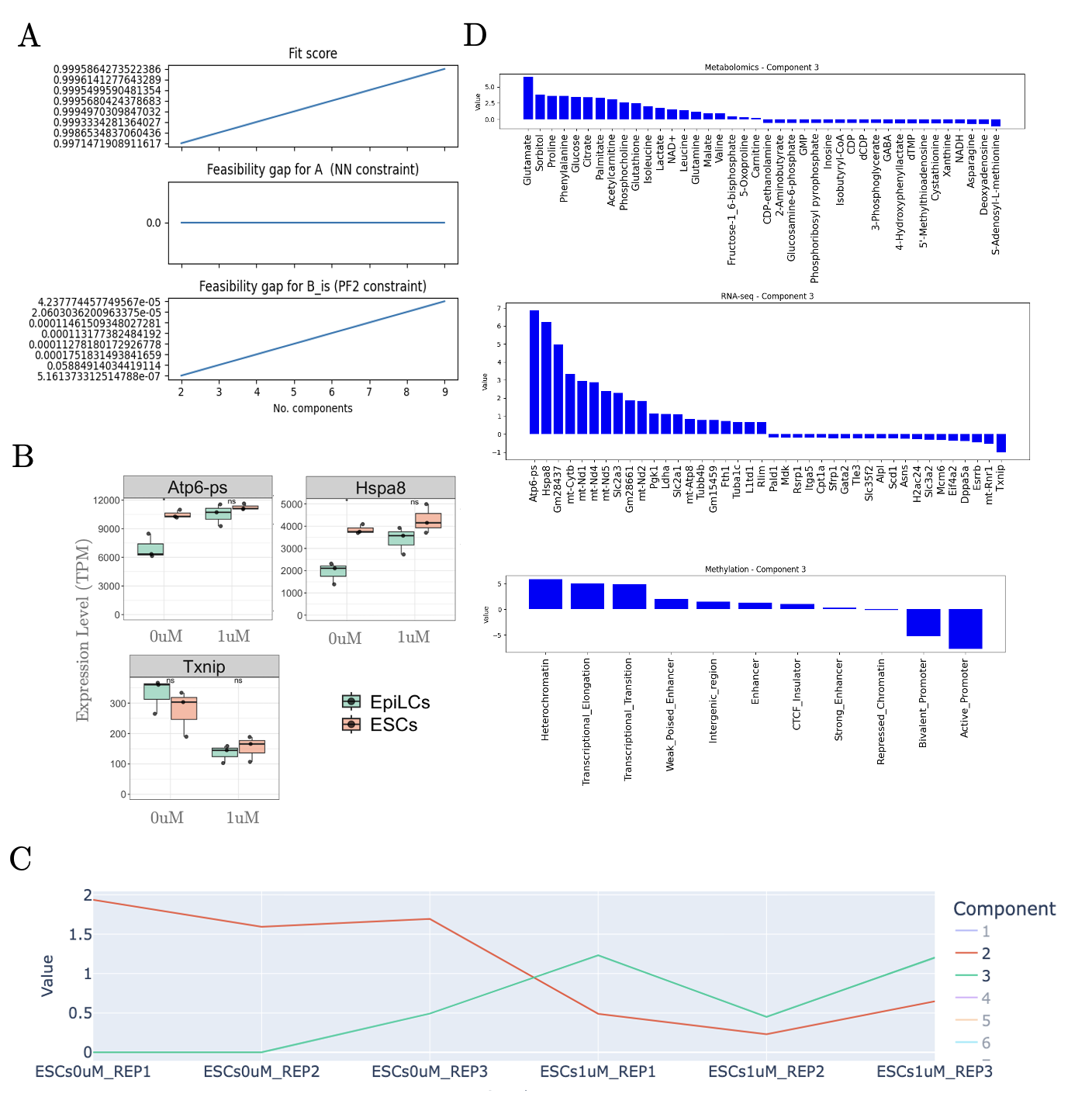}
      \caption{
         \textbf{Component based feature-level analysis identifies biologically relevant relevant arsenic response signatures in ESCs.} \textbf{A.} Model reproducibility analysis across independent initializations demonstrating stable component identification. \textbf{B.} Validation of key molecular targets: Expression of ATP synthase subunit (Atp6-ps), heat shock protein (Hspa8), and thioredoxin-interacting protein (Txnip) showing differential responses between ESCs and EpiLCs under arsenic treatment. \textbf{C.} Component score patterns for ESCs across replicates, showing consistent separation of control and treated samples in component space. Stars indicate the level of statistical significance based on a two-sample t-test. \textbf{D.} Top-ranking molecular features from each omics modality and component, demonstrating biological relevance of the computational approach.
    }
    \label{fig:feature-analysis}
\end{figure}

\section{Discussion}

\subsection{Epigenetic Mechanisms and Developmental Toxicity}
Our findings reveal critical insights into arsenic-induced epigenetic alterations in embryonic stem cells, with significant implications for developmental toxicity. The dominant weight-age of RRBS data in supports research demonstrating that DNA methylation changes represent among the earliest molecular events in arsenic toxicity [20,21]. This finding aligns with evidence that embryonic stem cells are particularly vulnerable to heavy metal exposure due to their reliance on epigenetic reprogramming for proper functioning [22].

 It has established that arsenic exposure during critical developmental windows can induce persistent epigenetic changes affecting stem cell function [23,24]. Heavy metals, including arsenic, are well-known modifiers of the epigenome, and stem cells rely heavily on epigenetic mechanisms for maintaining pluripotency and directing differentiation [22]. Our Component 2 pattern, which distinguished ESCs from EpiLCs regardless of treatment, suggests that arsenic may exploit existing epigenetic vulnerabilities in pluripotent cells.

The identification of global hypomethylation patterns in ESCs compared to EpiLCs provides mechanistic insight into arsenic toxicity. Recent studies have demonstrated that arsenic exposure induces global DNA hypomethylation through depletion of S-adenosylmethionine (SAM), the primary cellular methyl donor [25,26]. Our results suggest that ESCs may be more resilient to arsenic-induced hypomethylation due to their already hypomethylated state, representing a potential protective mechanism for maintaining pluripotency under environmental stress.

\subsection{Chromatin State-Specific Responses and Regulatory Mechanisms}
The ChromHMM-based aggregation of methylation data into 12 chromatin states provides insight into how arsenic affects specific regulatory elements. Research using ChIP-seq analysis has demonstrated that arsenic exposure affects genome-wide alterations of histone methylation profiles, particularly H3K4me3 and H3K27me3 modifications [27,28]. These modifications play crucial roles in gene regulation, with H3K4me3 associated with active gene expression and H3K27me3 linked to gene repression.

The ChromHMM annotation framework reveals distinct chromatin states with characteristic histone modification patterns that provide mechanistic insight into arsenic-induced epigenetic changes (Supplemental Figure S1). Our feature level component analysis (Figures 2D and 3D) reveals specific chromatin state responses that differ within celltype treatment conditions and between celltypees.

In ESCs (Figure 2C), component 2 shows strong activation patterns associated with intergenic regions (State 2) and active heterochromatin (State 3, characterizedd by H3K9me3), indicating that arsenic preferentially affects these regulatory elements in pluripotent cells. The activation of strong enhancers (State 8) is particularly pronounced in EpiLCs (Figure 2C), suggesting that differentiated cells rely more heavily on enhancer-mediated gene regulation that becomes vulnerable to arsenic disruption. Heterochromatin states show differential responses between cell types, with ESCs maintaining more stable repressive chromatin domains compared to EpiLCs.

Within the ESCs, arsenic treatment-specific perturbations to the epigenome, transcriptome, and metabolomics observed in Figures 3D demonstrate that active promoter regions (State 7, characterized by H3K4me3 and H3K27ac marks) and bivalent promoter (State 6, marked by H3K4me3, H3K27me3, and H3K9ac) show the most significant activation responses to arsenic exposure. In ESCs, the heterochromatin state and transcriptional elongation (State 3 and 10, marked with H3K9me3 and H3K36me3) show high levels of activation in unperturbed states, reflecting the vulnerability of developmentally important genes that maintain both activating and repressive marks in pluripotent cells.

The dynamic chromatin state sensitivity provides an explanation for the cell type-specific toxicity patterns. ESCs show preferential activation of intergenic regulatory elements and heterochromatin, consistent with their need to maintain transcriptional plasticity during environmental stress. EpiLCs demonstrate stronger enhancer activation, reflecting their reliance on tissue-specific enhancer networks that become disrupted during arsenic exposure. This pattern suggests that arsenic toxicity mechanisms differ fundamentally between pluripotent and differentiated cell states.

\subsection{Metabolic Disruption and Cellular Stress Responses}
The identification of key metabolic biomarkers provides biochemical validation of arsenic-induced cellular stress. The presence of 5-oxoproline among highly variable metabolites confirms with existing literature that this compound serves as a sensitive biomarker for arsenic exposure and glutathione depletion [30,31]. Research has established that 5-oxoproline levels increase during glutathione recycling under oxidative stress conditions, making it a valuable indicator of arsenic-induced cellular damage [32].

Our analysis also revealed significant perturbations in S-adenosyl-L-methionine metabolism, which directly connects to the epigenetic effects observed in component 3 of the within cell type feature analysis for the ESCs. Arsenic's competition for SAM-provided methyl groups creates a metabolic bottleneck that affects multiple cellular processes, including DNA, RNA, and histone methylation [33]. This metabolic-epigenetic axis represents a critical mechanism through which arsenic exerts its toxic effects on developing cells.

The identification of mitochondrial dysfunction markers, including ATP synthase subunit changes, aligns with research demonstrating that arsenic induces metabolic reprogramming in stem cells. Studies have shown that arsenic exposure can shift cellular metabolism toward glycolysis and away from oxidative phosphorylation, potentially as an adaptive response to mitochondrial damage [34].

\subsection{Gene Expression Networks and Stress Response Pathways}
The high variability of Gapdh in our factor analysis provides insight into arsenic detoxification mechanisms. Research has established that Gapdh functions as a fortuitous arsenate reductase, catalyzing the reduction of arsenate (As\textsuperscript{V}) to arsenite (As\textsuperscript{III}) [35,36]. The newly discovered Gapdh-ArsJ pathway provides arsenate resistance through formation of 1-arseno-3-phosphoglycerate, which is then extruded from cells [37]. Our findings suggest that cells may modulate Gapdh expression as part of their arsenate detoxification strategy.

The differential expression of Dnmt3b between ESCs and EpiLCs under arsenic exposure supports the hypothesis that arsenic disrupts normal methylation programming during early development [38]. Dnmt3b is essential for de novo methylation during embryonic development, and its dysregulation could have profound implications for developmental toxicity [39]. Previous studies have linked arsenic exposure to altered Dnmt3b function and global hypomethylation patterns consistent with our observations [40].

Heat shock protein (Hspa8) identification among key variable genes reflects the cellular stress response to arsenic exposure. HSP proteins play critical roles in protein folding and cellular stress management, and their upregulation is a hallmark of arsenic-induced cellular stress [41]. The coordinated response of multiple stress-related genes in our analysis suggests that arsenic exposure triggers comprehensive cellular stress response pathways in both ESCs and EpiLCs.

\subsection{Implications for Developmental Toxicology and Risk Assessment}
Our findings have significant implications for understanding arsenic's developmental toxicity and improving risk assessment methodologies. The three-component structure identified by our analysis provides a framework for understanding arsenic toxicity at the systems level.

Recent research has demonstrated that prenatal arsenic exposure can induce transgenerational epigenetic effects, with alterations in histone modifications persisting across generations [42,43]. Our component-based analysis framework could be extended to study these transgenerational effects, potentially identifying epigenetic signatures that predict long-term developmental consequences.

The identification of chromatin state-specific responses provides new targets for biomarker development in environmental health. Rather than monitoring individual genes or metabolites, our approach suggests that chromatin state-aggregated methylation patterns could serve as more robust indicators of arsenic exposure and its biological effects [44]. This could improve early detection of arsenic-induced developmental abnormalities and guide intervention strategies.

\subsection{Computational Advances and Future Directions}
This study demonstrates that coupled matrix factorization can effectively integrate heterogeneous molecular data to reveal coordinated responses to environmental exposures. The approach shows particular promise for regulatory toxicology applications where comprehensive molecular characterization is needed for chemical risk assessment. The integration of multi-omics data through coupled matrix factorization represents a promising avenue for advancing computational toxicology and environmental health research.

Future applications could extend this approach to other environmental toxicants and complex exposure scenarios, contributing to the development of next-generation toxicity testing strategies that reduce reliance on animal testing while improving predictive accuracy for human health risk assessment [45]. The framework could also be adapted for dose-response modeling and mixture toxicity assessment, critical areas for environmental health research.

The success of our ChromHMM-based aggregation strategy suggests that incorporating additional layers of regulatory annotation could further enhance the biological interpretability of multi-omics integration approaches. Future work could integrate additional epigenetic marks, chromatin accessibility data, and three-dimensional chromatin structure information to provide even more comprehensive views of arsenic-induced toxicity mechanisms.

\section{Supplemental Material}

\subsection{Supplemental Figures}

\begin{figure}[H]
    \centering
    \includegraphics[width=0.8\linewidth]{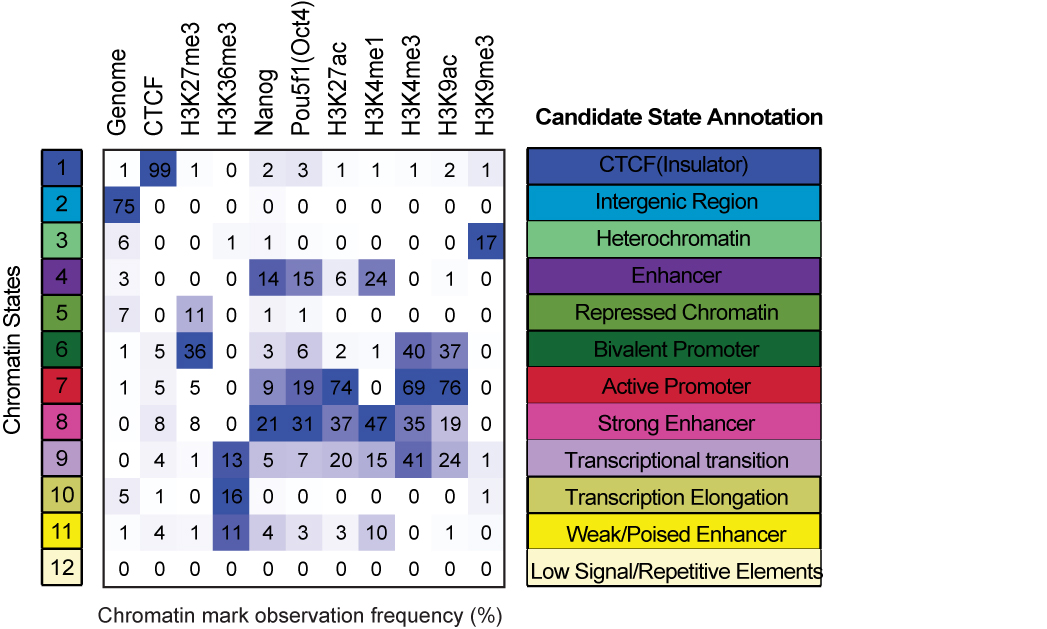}
      \caption{
        \textbf{Supplemental Figure S1: ChromHMM chromatin state annotations and histone modification patterns.} The heatmap shows the chromatin mark observation frequency (\%) for 12 candidate chromatin states used in the analysis. Each state is characterized by distinct combinations of histone modifications and transcription factor binding (CTCF). State 1 represents CTCF insulators with 99\% CTCF occupancy, state 2 shows intergenic regions with 75\% frequency, States 6-8 represent different promoter and enhancer categories with varying H3K4me3, H3K27ac, and H3K27me3 patterns, and states 3-5 represent repressed chromatin domains. The bivalent promoter state (State 6) shows co-occurrence of both activating (H3K4me3) and repressive (H3K27me3) marks, characteristic of poised developmental genes in embryonic stem cells. State 7 (active promoters) and state 8 (strong enhancers) show high frequencies of activating marks, making them particularly vulnerable to arsenic-induced disruption. State 9 represents transcriptional transition regions with moderate levels of multiple histone marks, while States 11-12 show weak enhancer signals or low signal regions respectively.
    }
    \label{fig:supplemental-chromatin-states}
\end{figure}

\section{References}

\begin{enumerate}
    \item Mohammed Abdul, K. S., Jayasinghe, S. S., Chandana, E. P. S., Jayasumana, C., \& De Silva, P. M. C. S. (2015). Arsenic and human health effects: A review. \textit{Environmental Toxicology and Pharmacology}, \textit{40}(3), 828--846.
    
    \item Gasser, M., Lenglet, S., Bararpour, N., et al. (2023). Arsenic induces metabolome remodeling in mature human adipocytes. \textit{Toxicology}, \textit{500}, 153672.
    
    \item Hughes, M. F., Beck, B. D., Chen, Y., Lewis, A. S., \& Thomas, D. J. (2011). Arsenic exposure and toxicology: a historical perspective. \textit{Toxicological Sciences}, \textit{123}(2), 305--332.
    
    \item Roald, M. (2023). MatCoupLy: Learning coupled matrix factorizations with Python. \textit{SoftwareX}, \textit{21}, 101292.
    
    \item Pierre-Jean, M., Deleuze, J. F., Le Floch, E., \& Mauger, F. (2020). Clustering and variable selection evaluation of 13 unsupervised methods for multi-omics data integration. \textit{Briefings in Bioinformatics}, \textit{21}(6), 2011--2030.
    
    \item Singh, A., Shannon, C. P., Gautier, B., et al. (2019). DIABLO: an integrative approach for identifying key molecular drivers from multi-omics assays. \textit{Bioinformatics}, \textit{35}(17), 3055--3062.
    
    \item Argelaguet, R., Velten, B., Arnol, D., et al. (2018). Multi-Omics Factor Analysis—a framework for unsupervised integration of multi-omics data sets. \textit{Molecular Systems Biology}, \textit{14}(6), e8124.
    
    \item Welch, J. D., Kozareva, V., Ferreira, A., et al. (2019). Single-cell multi-omic integration compares and contrasts features of brain cell identity. \textit{Cell}, \textit{177}(7), 1873--1887.
    
    \item Cao, K., \& Gong, X. (2022). Multi-omics single-cell data integration and regulatory inference with graph-linked embedding. \textit{Nature Biotechnology}, \textit{40}(10), 1458--1466.
    
    \item Bagherian, M., Kim, R. B., Jiang, C., Sartor, M. A., Derksen, H., \& Najarian, K. (2021). Coupled matrix--matrix and coupled tensor--matrix completion methods for predicting drug--target interactions. \textit{Briefings in Bioinformatics}, \textit{22}(2), 2161--2171.
    
    \item Erbe, R., Kessler, M. D., Favorov, A. V., Easwaran, H., Gaykalova, D. A., \& Fertig, E. J. (2020). Matrix factorization and transfer learning uncover regulatory biology across multiple single-cell ATAC-seq data sets. \textit{Nucleic Acids Research}, \textit{48}(12), e68.
    
    \item Duren, Z., Chen, X., Zamanighomi, M., et al. (2018). Integrative analysis of single-cell genomics data by coupled nonnegative matrix factorizations. \textit{Proceedings of the National Academy of Sciences}, \textit{115}(30), 7723--7728.
    
    \item Yang, J., Williams, A. E., Petre, B. J., et al. (2024). Integrative, high-resolution analysis of single cell gene expression across experimental conditions with PARAFAC2-RISE. \textit{bioRxiv}.
    
    \item Zhang, Z., Shen, J., Song, Z., et al. (2024). Advances in cancer DNA methylation analysis with methPLIER: use of non-negative matrix factorization and knowledge-based constraints to enhance biological interpretability. \textit{Experimental \& Molecular Medicine}, \textit{56}, 394--405.
    
    \item Huang, S., Wang, Z., Zhang, Y., et al. (2024). FedPAR: Federated PARAFAC2 tensor factorization for computational phenotyping. \textit{IISE Transactions on Healthcare Systems Engineering}, \textit{14}(3), 185--197.
    
    \item Roald, M., Schenker, C., Calhoun, V. D., Adali, T., Bro, R., Cohen, J. E., \& Acar, E. (2024). PARAFAC2-based Coupled Matrix and Tensor Factorizations with Constraints. arXiv preprint arXiv:2406.12338.
    
    \item Acar, E., Papalexakis, E. E., Gürdeniz, G., et al. (2014). Structure-revealing data fusion. \textit{BMC Bioinformatics}, \textit{15}, 239.
    
    \item Harshman, R. A. (1972). PARAFAC2: Mathematical and technical notes. \textit{UCLA Working Papers in Phonetics}, \textit{22}, 30--44.
    
    \item Wei, G. ChromHMM inferred annotations for mouse embryonic stem cells (mm10). GitHub repository. \url{https://github.com/guifengwei/ChromHMM\_mESC\_mm10}
    
    \item Ren, X., Kuan, P. F. (2019). methylGSA: a Bioconductor package and Shiny app for DNA methylation data length bias adjustment in gene set testing. \textit{Bioinformatics}, \textit{35}(11), 1958--1959.
    
    \item Smith, L. N., \& Welch, J. D. (2021). Environmental epigenetics and a unified theory of the molecular aspects of evolution: A neo-Lamarckian concept that facilitates neo-Darwinian evolution. \textit{Environmental Epigenetics}, \textit{7}(1), dvab005.
    
    \item Mokra, K., Konopacka, M., \& Olichwier, A. (2024). Epigenetic toxicity of heavy metals -- implications for embryonic stem cells. \textit{Science of The Total Environment}, \textit{955}, 176706.
    
    \item Tyler, C. R., \& Allan, A. M. (2014). The effects of arsenic exposure on neurological and cognitive dysfunction in human and rodent studies: a review. \textit{Current Environmental Health Reports}, \textit{1}(2), 132--147.
    
    \item Rager, J. E., Auerbach, S. S., Chappell, G. A., et al. (2017). Benchmark dose modeling estimates of the concentrations of inorganic arsenic that induce changes to the neonatal transcriptome, proteome, and epigenome in a pregnancy cohort. \textit{Chemical Research in Toxicology}, \textit{30}(10), 1911--1920.
    
    \item Bailey, K. A., \& Fry, R. C. (2014). Long-term health consequences of prenatal arsenic exposure: links to prevention and treatment. \textit{Advances in Experimental Medicine and Biology}, \textit{827}, 333--347.
    
    \item Reichard, J. F., Schnekenburger, M., \& Puga, A. (2007). Long term low-dose arsenic exposure induces loss of DNA methylation. \textit{Biochemical and Biophysical Research Communications}, \textit{352}(1), 188--192.
    
    \item Chervona, Y., \& Costa, M. (2012). Histone modifications and cancer: biomarkers of prognosis? \textit{American Journal of Cancer Research}, \textit{2}(5), 589--597.
    
    \item Mo, J., Xia, Y., Wade, T. J., et al. (2022). Genome-wide alteration of histone methylation profiles associated with cognitive changes in response to developmental arsenic exposure in mice. \textit{Computational and Structural Biotechnology Journal}, \textit{20}, 1133--1145.
    
    \item Cronican, A. A., Fitz, N. F., Carter, A., et al. (2013). Genome-wide alteration of histone H3K9me3 binding sites and gene expression profiles in the mouse hippocampus after chronic arsenic exposure. \textit{Toxicological Sciences}, \textit{135}(2), 405--414.
    
    \item Kumar, A., \& Xagoraraki, I. (2010). Pharmaceuticals, personal care products and endocrine-disrupting chemicals in U.S. surface and finished drinking waters. \textit{Journal of Water and Health}, \textit{8}(4), 674--695.
    
    \item Mazumder, D. N. (2008). Chronic arsenic toxicity \& human health. \textit{Indian Journal of Medical Research}, \textit{128}(4), 436--447.
    
    \item Smeester, L., Rager, J. E., Bailey, K. A., et al. (2011). Epigenetic changes in individuals with arsenicosis. \textit{Chemical Research in Toxicology}, \textit{24}(2), 165--167.
    
    \item Hughes, M. F. (2002). Arsenic toxicity and potential mechanisms of action. \textit{Toxicology Letters}, \textit{133}(1), 1--16.
    
    \item Tokar, E. J., Diwan, B. A., Ward, J. M., Delker, D. A., \& Waalkes, M. P. (2011). Carcinogenic effects of "whole life" exposure to inorganic arsenic in CD1 mice. \textit{Toxicological Sciences}, \textit{119}(1), 73--83.
    
    \item Némethi, B., Csanaky, I., \& Gregus, Z. (2006). Effect of an inactivator of glyceraldehyde-3-phosphate dehydrogenase, a fortuitous arsenate reductase, on disposition of arsenate in rats. \textit{Toxicological Sciences}, \textit{90}(1), 49--60.
    
    \item Gregus, Z., \& Németi, B. (2005). The glycolytic enzyme glyceraldehyde-3-phosphate dehydrogenase works as an arsenate reductase in human red blood cells and rat liver cytosol. \textit{Toxicological Sciences}, \textit{85}(2), 859--869.
    
    \item Ye, J., Chang, Y., Liu, X., et al. (2016). Synergistic interaction of glyceraldehydes-3-phosphate dehydrogenase and ArsJ, a novel organoarsenical efflux permease, confers arsenate resistance. \textit{Applied and Environmental Microbiology}, \textit{82}(20), 6090--6103.
    
    \item Okano, M., Bell, D. W., Haber, D. A., \& Li, E. (1999). DNA methyltransferases Dnmt3a and Dnmt3b are essential for de novo methylation and mammalian development. \textit{Cell}, \textit{99}(3), 247--257.
    
    \item Xu, W., Xu, M., Wang, L., et al. (2019). Integrative analysis of DNA methylation and gene expression identified cervical cancer-specific diagnostic biomarkers. \textit{Signal Transduction and Targeted Therapy}, \textit{4}, 55.
    
    \item Pecori, F., Yokota, I., Hanamatsu, H., et al. (2021). A defined glycosylation regulatory network modulates total glycome dynamics during pluripotency state transition. \textit{Scientific Reports}, \textit{11}, 1276.
    
    \item Huang, C., Ke, Q., Costa, M., \& Shi, X. (2004). Molecular mechanisms of arsenic carcinogenesis. \textit{Molecular and Cellular Biochemistry}, \textit{255}(1-2), 57--66.
    
    \item Li, J., Goyer, R. A., \& Waalkes, M. P. (2003). Toxicity of inorganic arsenic. In \textit{Handbook on the Toxicology of Metals} (pp. 321--338). Elsevier.
    
    \item Tyler, C. R., Hafez, A. K., Echeverria, V., et al. (2018). Developmental exposure to 50 parts-per-billion arsenic influences histone modifications and associated epigenetic machinery in a region- and sex-specific manner in the adult mouse brain. \textit{Toxicology and Applied Pharmacology}, \textit{288}, 40--51.
    
    \item Cantini, L., Zakeri, P., Hernandez, C., et al. (2021). Benchmarking joint multi-omics dimensionality reduction approaches for the study of cancer. \textit{Nature Communications}, \textit{12}, 124.
    
    \item Thomas, R. S., Paules, R. S., Simeonov, A., et al. (2018). The US Federal Tox21 Program: A strategic and operational plan for continued leadership. \textit{ALTEX-Alternatives to animal experimentation}, \textit{35}(2), 163--168.
    
\end{enumerate}

\end{document}